\documentstyle[osa,aps,prb,epsfig]{revtex}
\begin{document}
\twocolumn[\hsize\textwidth\columnwidth\hsize\csname
@twocolumnfalse\endcsname
\title{
Explanation of small I$_c$R$_n$ values observed in inhomogeneous
superconductors}
\author{Hyun-Tak Kim (htkim@etri.re.kr, ~kimht45@hotmail.com)}
\address{Telecom. Basic Research Lab., ETRI, Daejeon 305-350, Korea}
\maketitle{}
\begin{abstract}
For an inhomogeneous high-$T_c$ superconductor, band-filling
dependence of Josephson I$_c$R$_n$ product is deduced at $T$=0 K
by means of measurement; this is an extension of the Ambegaokar -
Baratoff (AB) theory based on the $s$-wave theory. The product is
given by $J_{obs}R_n
{\equiv}\frac{\pi}{2}\rho\triangle_i={\rho}J_iR_n$, where
0$<\rho\le$1 is band filling (or local density), and $\triangle_i$
is the intrinsic superconducting true gap and small. When
$\rho$=1, $J_{obs}R_n=J_iR_n$ is the intrinsic Josephson true
product (or the AB product), where $J_i$ is the intrinsic
Josephson true current occurring by Cooper pair. When 0$<\rho<$1,
$J_{obs}R_n$ is an average of $J_iR_n$ over the measurement region
and is the effect of measurement. The $J_{obs}R_n$ explains small
I$_c$R$_n$ values observed by experiments. Furthermore, the
intrinsic gap, 8.5$<{\triangle_i}<$17 meV, is analyzed from
I$_c$R$_n$ data of Bi$_2$Sr$_2$CaCu$_2$O$_{8+x}$. $d$-wave
superconductive components do not exist in Bi-2212 crystals.
\end{abstract}
]
From the discovery of a high-$T_c$  superconductor until recently,
pairing symmetry for the mechanism of high-$T_c$ superconductivity
has been controversial because intrinsic physical information of
superconductors is  not obtained for intrinsically inhomogeneous
superconductors$^{1-5}$ with a metal phase and an insulator phase
with the $d_{x^2-y^2}$-wave symmetry.$^6$ The intrinsic
inhomogeneity in which a homogeneous metal region is about 14$\AA$
was revealed by scanning tunnelling microscopy.$^1$ The
inhomogeneity is due to the metal-insulator instability.$^5$ In
inhomogeneous superconductors, the fact that the energy gap
decreases with an increasing local density was also revealed by
experiments$^{1-4}$ and a theoretical consideration$^6$. Recently,
for an inhomogeneous superconductor, an analysis method for the
intrinsic density of states$^{7,8}$ and the intrinsic
superconducting gap$^6$ was developed by means of measurement. The
method disclosed the identity of gap anisotropy and revealed that
pairing symmetry of a high-$T_c$ superconductor is an
$s$-wave.$^6$

However, there  are  still two unresolved problems to be clarified
on the Josephson I$_c$R$_n$ product. One is that the product
decreases with an increasing superconducting gap (Fig.
1).$^{9,10}$ The other is that I$_c$R$_n$ values in c-axis
Josephson pair (or intrinsic Josephson) tunneling experiments are
much smaller than the all-$s$-wave Ambegaokar-Baratoff
limit.$^{11-19}$ On the basis of this experimental result, it has
been interpreted that the ratio of the $s$-wave component to the
full $d$-wave one in the $s+d$ mixed states is very small. In
addition, the smaller I$_c$R$_n$ value was also observed in
Pb/I/NbSe$_2$ junctions.$^{21}$

In this paper, we deduce band-filling (or doping) dependence of
the Josephson I$_c$R$_n$ product, by using the means of
measurement suggested in previous papers$^{6,7}$; this is an
extension of the Ambegaokar - Baratoff theory$^{20}$. The
intrinsic superconducting gap is analyzed from early published
I$_c$R$_n$ data.

Fractional charge has been demonstrated in previous
papers.$^{6,7}$ These will be reviewed briefly. In an
inhomogeneous superconductor with two phases of a metal region and
an insulating region, when it is measured such

\begin{figure}
\vspace{-2.0cm}
\centerline{\epsfysize=8.0cm\epsfxsize=7cm\epsfbox{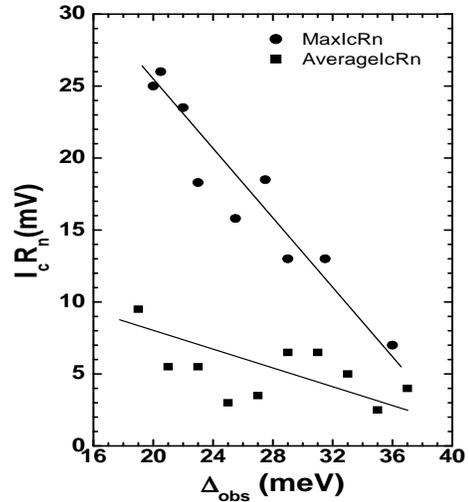}}
\vspace{0.0cm} \caption{Data was cited from A. Mourachkine, J.
Supercond. 14 (2001) 375.$^9$ Maximum and average $I_cR_n$
products vs energy gap, which measured in 110
Bi$_2$Sr$_2$CaCu$_2$O$_{8+x}$ break junctions. The $I_cR_n$
product decreases with an increase of the observed energy gap,
which differs from $I_cR_n\propto\triangle$ in the
Ambegaokar-Baratoff theory. In addition, linear lines are a trend
not fittings.}
\end{figure}

as photoemission spectroscopy, a spectral-weight value in
$k$-space is observed, but the inhomogeneous phases are not able
to be deduced from the observed spectral weight. In other words, a
reverse transformation from $k$-space into $real$-space is not
defined, (Fig. 2 (a)). This indicates that two real- and $k$-
spaces are not mathematically equivalent. The inhomogeneous
superconductor is different from the metal with both the
electronic structure of $one~electron~per~atom$ and mathematically
equivalence between two spaces. In order to overcome this problem,
we think out that an measured data is an averaged data. When the
inhomogeneous superconductor is measured, carriers in the metal
region should be averaged over lattices (or atoms) in the entire
measurement region. Then, the inhomogeneous superconductor is
changed into a homogeneous one with the electronic structure of
$one~effective~charge~per~atom$, (Fig. 2 (b)). The observed
effective charge becomes $e'={\rho}e$, where 0$<{\rho}=n/L{\le}$1
is band filling (or local density), $n$ is the number of carriers
in the metal region, and $L$ is the number of total lattices in
the measurement region. The fractional effective charge is
justified only when the inhomogeneous system is measured.
Otherwise, it becomes true charge in the metal region.

\begin{figure}
\vspace{-0.0cm}
\centerline{\epsfysize=9.0cm\epsfxsize=8.4cm\epsfbox{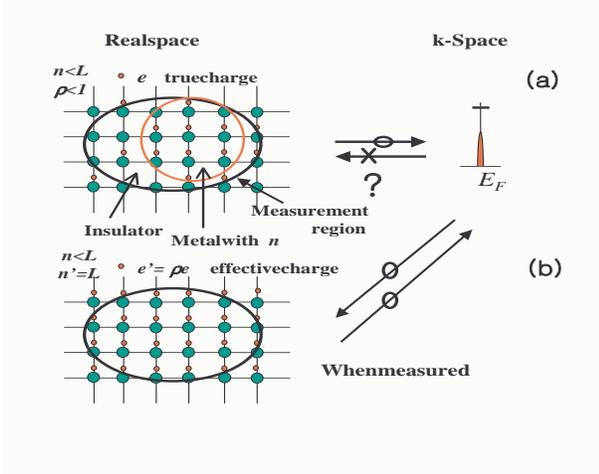}}
\vspace{-2.0cm} \caption{(a) In an inhomogeneous superconductor, a
reverse transformation from $k$-space to $real$-space is not
defined, which is a problem. $n$ is the number of electrons in
metal region. $L$ is the number of lattices in the measurement
region. 0$<\rho=\frac{n}{L}\le$1 is defined. The spectral wight in
$k$-space increases with an increasing $\rho$. (b) When the metal
region is averaged over lattices in the measurement region, the
inhomogeneous superconductor become homogeneous when measured. The
two spaces are mathematically equivalent. $e'={\rho}e$ is a
fractional effective charge. When $\rho$=1, it is metal.}
\end{figure}

When the concept of measurement is applied to an inhomogeneous
superconductor, the observed energy gap, $\triangle_{obs}$, was
given by

\begin{eqnarray}
{\triangle}_{obs}={\triangle}_i/{\rho} ,
\end{eqnarray}
where $\triangle_i$ is the intrinsic  superconducting true gap
determined by the minimum bias voltage. For understanding of Eq.
(1), Fig. 3 is given. The 0$<{\rho}{\le}$1 is band filling (local
density or homogeneous factor), which indicates the extent of the
metal region. The validity of Eq. (1) was given by many tunneling
experiments.$^6$

Ambegaokar and Baratoff$^{20}$  generalized the Josephson tunnel
theory and calculated the coherent tunnelling supercurrent on the
basis of the BCS theory for a $s$-wave homogeneous superconductor.
The suppercurrent at $T$=0 K was given by

\begin{eqnarray}
J = \frac{\pi}{2}R_n^{-1}{\triangle},
\end{eqnarray}

where $R_n = (2{\pi}h/e^2T)$, $T$ is the tunneling matrix, and
$\triangle$ is a superconducting energy gap.

\begin{figure}
\vspace{-1.0cm}
\centerline{\epsfysize=9.0cm\epsfxsize=8.4cm\epsfbox{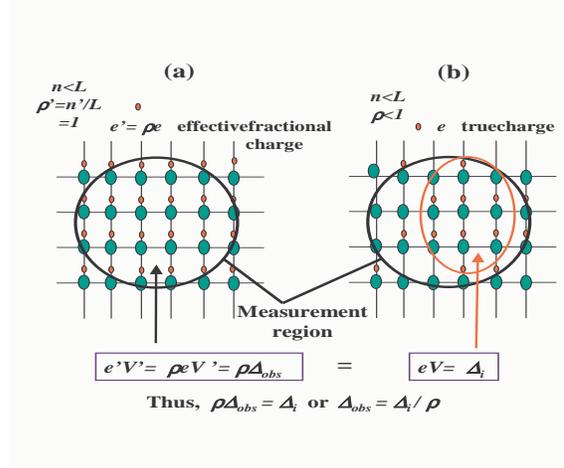}}
\vspace{-2.0cm} \caption{(a) When an inhomogeneous superconductor
is measured, the homogeneous metal region in Fig. 2 (b) is
averaged over lattices in the measurement region. Then,
$e'V'={\rho}eV'={\rho}\triangle_{obs}$ is given. (b) In an
inhomogeneous superconductor, if only the homogeneous metal region
is measured, $eV=\triangle_i$ is given.}
\end{figure}

In an inhomogeneous superconductor, the averaged metallic system
has the electronic structure of one effective charge per atom, as
shown in Fig. 2 (b), which is mathematically equivalent to the
electronic structure of the metal used in the BCS theory. The
metal for $k$-space used in the BCS theory has the electronic
structure of one electron per atom, as shown when $\rho$=1 in Fig.
2 (b). The Josephson current and the product derived in the
Ambegaokar and Baratoff theory can be used without formula's
change even in the inhomogeneous superconductor by replacing true
charge by the effective charge because the averaged effective
charge is invariant under transformation. Particular calculations
are not necessary because it had already been given by Ambegaokar
and Baratoff$^{20}$. In addition, similar calculations have been
given when the Brinkman-Rice picture was extended$^7$. Thus, the
observed supercurrent, $J_{obs}$, is given by substituting $e$ in
$R_n$ and $\triangle$ with $e'={\rho}e$ and
$\triangle_{obs}$=${\triangle_i}/{\rho}$ by
\begin{eqnarray}
J_{obs} {\equiv}(\frac{e^2T}{4h}){\rho}{\triangle_i}={\rho}J_i,
\end{eqnarray}
where $J_i$ is the intrinsic true supercurrent. The observed
Josephson product is also given by Eq. (3) by

\begin{eqnarray}
J_{obs}R_n
 {\equiv} \frac{\pi}{2}{\rho}^2{\triangle_{obs}}{\equiv}\frac{\pi}{2}{\rho}{\triangle_i}={\rho}J_iR_n,
\end{eqnarray}

where $\triangle_i$ is constant.

When ${\rho}$ =1, Eqs (3) and (4) are the intrinsic supercurrent
and the intrinsic product (or Ambegaokar and Baratoff product)
caused by true Cooper pairs, respectively. When 0$<{\rho}<$1, the
equations correspond to the averages of the intrinsic supercurrent
and the intrinsic product over the measurement region and are the
effect of measurement. Eqs (3) and (4) are the extended Ambegaokar
- Baratoff (AB) Josephson current and product, respectively.

\begin{figure}
\vspace{-0.5cm}
\centerline{\epsfysize=8.0cm\epsfxsize=7cm\epsfbox{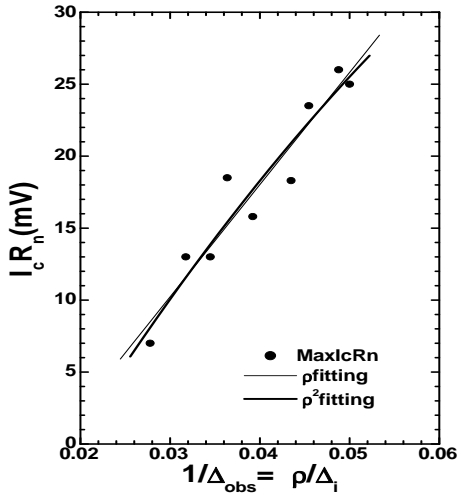}}
\vspace{0.0cm} \caption{This shows fittings of the maximum
$I_cR_n$ data in Fig. 1. An inverse energy gap as x-axis, on the
basis of Eq. (1), is taken to fit the $\rho$ dependence. Even
though it is not easy to distinguish a more good fitting between
linear and square fittings with respect to $\rho$, it can be
interpreted that the $\rho^2$ fitting of Eq. (4) is more
satisfied.}
\end{figure}

The Josephson product of Eq. (4), which decrease with decreasing
${\rho}$, fits well the experimental data measured by break
junctions (Fig. 4). This reveals that, at a fixed doping level,
the observed Josephson product decreases with an increasing energy
gap $^{9,10}$,
 and that Bi$_2$Sr$_2$CaCu$_2$O$_{8+x}$ (Bi-2212) crystals are
 inhomogeneous, which is a general characteristic of high-$T_c$ superconductors$^{1-5}$ and
 the reason why the AB product is not applied directly to
 experimental data. Moreover, because Eq. (4) with $\rho\ne$1 is an average of the
true $I_cR_n$ product value (or the AB product value), which is
observed in only the metal region with $\rho$=1 in Fig. 2 (a),
$d$-wave superconductive components do not exist in the
homogeneous metal region in Bi-2212 crystals. Note that
Mourachkine$^{10}$ discussed that the decrease of the product in
Fig. (1) is not intrinsic effect and due to inhomogeneity.

 The ${\rho}$ dependence in Eq. (4) comes from Eq. (3), which agrees
with a result observed by the intrinsic Josephson junction.$^{16}$
Note that the magnitude of the intrinsic AB product is basically
very small because ${\triangle_i}$ is small. Considering that the
anisotropy of the number of carriers in the c-axis and ab-plane is
large (${\rho_{c-axis}}<<{\rho_{ab-plane}}$), the product observed
in the c-axis is naturally much less than that in the ab-plane.
Thus, the observed small I$_c$R$_n$ values$^{10-18,20}$ can be
explained by Eq. (4). Additionally, for a Josephson junction by
two superconductors with different energy gaps, the I$_c$R$_n$
product derived by Anderson$^{22}$ is in the context of the above
analysis.

We analyze the intrinsic gap of Bi-2212 from experimental data,
using Eq. (4). Irie $et~al.^{16}$ suggested that
$I_cR_n\approx$13.3 meV observed by the intrinsic Josephson
junction is $\frac{1}{3}$ of $I_cR_n\approx$ 40 meV using
$\triangle_{obs}\approx$25 meV and Eq. (4) with $\rho$=1. The true
product value is much less than 40 meV, when both $\rho\ne$1 and
$\triangle_{obs}\ne\triangle_i$ are considered. The intrinsic gap,
$\triangle_i\approx$8.5 meV, is obtained from the observed
$I_cR_n\approx$13.3 meV by Eq. (4) with $\rho$=1. The intrinsic
true gap is slightly larger than 8.5 meV because $\rho<$1
slightly. Mourachkine$^{9,10}$ observed the maximum Josephson
product of $I_cR_n\approx$ 26 meV for an over-doped crystal, which
can be regarded as $\rho\approx$1 without a pseudogap$^{23}$, at
the minimum energy gap of $\triangle_{obs}\approx$21 meV. The
intrinsic true gap, $\triangle_i\approx$~16.5 meV, is obtained by
Eq. (4) with $\rho$=1. The intrinsic true gap is less than that
analyzed by Mourachkine. Thus, we conclude that the analyzed
intrinsic gap is in 8.5 $<{\triangle_i}<$17 meV.

In conclusion, for inhomogeneous high-$T_c$ superconductors,
without the $d$-wave theory, Eq. (4) based on the $s$-wave theory
explains the small Josephson-product values observed by
experiments.

\section*{ACKNOWLEDGEMENTS}
We acknowledges R. A. Klemm for very valuable comments and A.
Mourachkine for the use of Fig. 1 and very valuable comments.

\end{document}